\documentclass[aps,prl,twocolumn,showpacs]{revtex4}

\usepackage{amsmath}
\usepackage{amssymb}
\usepackage{graphpap}
\usepackage[dvips]{graphicx}
\usepackage[dvips]{graphics}
\usepackage{color}
\long\def\@makesidecaption#1#2{%
   \parbox[b]{\@tempdima}{\captionstyle{\floatlegendstyle
                                         #1\floatcounterend}#2}}

\begin{document}

% Title of the article
\title{Graphene quantum dots in perpendicular magnetic fields}

% Abbreviated title for the page headers
%\titlerunning{Graphene quantum dots}

% Authors
\author{%
% First Author\textsuperscript{\Ast,\textsf{\bfseries 1}},
%  Second Author\textsuperscript{\textsf{\bfseries 2}},
%  Third Author\textsuperscript{\textsf{\bfseries 3}}, \ldots
J. G\"uttinger, C. Stampfer, T. Frey, T. Ihn and K. Ensslin}
 \affiliation{Solid State Physics Laboratory, ETH Zurich, 8093 Zurich, Switzerland}
% Abbreviated list of authors for the page headers
%\authorrunning{J. G\"uttinger et al.}

%E-mail-address of corresponding author
%\mail{e-mail
%  \textsf{guettinj@phys.ethz.ch}, Phone:
%  +41-44-6332836 , Fax: +41-44-6331146 }

% author's affiliations/addresses
%\institute{%
%  %\textsuperscript{1}\,
%  Solid State Physics Laboratory, ETH Zurich, 8093 Zurich, Switzerland }
  %\\
  %\textsuperscript{2}\,Second address\\
  %\textsuperscript{3}\,Third address}

%\received{XXXX, revised XXXX, accepted XXXX} % do not change, will be filled in by the publisher
%\published{XXXX} % do not change, will be filled in by the publisher

%Please select four to six PACS-codes from the enclosed list (PACS.txt) or from www.aip.org/pacs)
 % For example: 71.20.Ps

%\abstract{%
% This is a macro for the typesetting of two-column text in an
% abstract. It will typeset the two arguments in \abstcol{}{} as the
% left and right column inside the abstract box. At the
% columnbreak there will be always a columnbreak (\par), so both
% columns start with a new paragraph. No automatic column height
% balancing is done.
%
% If used with a \titlefigure it will silently output both
% parameters as consecutive paragraphs.
%
% The macro is defined exclusively inside the argument of \abstract{};
% if used outside it will raise an error.
%
% Usage: \abstcol{<left column>}{<right column>}
%\abstcol{%

\begin{abstract}
  We report transport experiments on graphene quantum dots. We focus on excited state spectra in the near vicinity of the charge neutrality point and signatures of the electron-hole crossover as a function of a perpendicular magnetic field. Coulomb blockade resonances of a 50~nm wide and 80~nm long dot are visible at all gate voltages across the transport gap ranging from hole to electron transport. The magnetic field dependence of more than
40 states as a function of the back gate voltage can be interpreted in terms of the unique evolution of the diamagnetic spectrum of a graphene dot including the formation of the $E=0$ Landau level, situated in the center of the transport gap, and marking the electron-hole crossover. 
\end{abstract}
% The class file requires the standard graphicx Latex package. See the 'LaTeX
% standard graphics and color packages documentation' for more information at
% <http://tug.ctan.org/tex-archive/macros/latex/required/graphics/grfguide.pdf>.
%
% Accepted figure file formats depend on which LaTeX flavour is used.
% Classic LaTeX is always able to use Encapsulted Postscript (EPS);
% PDFLaTeX can't use this but accepts PDF, JPG, PNG, and GIF formats.
%
% See examples for implementing graphics in floating figure environments later in this file.
% If \titlefigure is given, it takes as its mandatory parameter the
% name (without extension) of some figure file.
%\titlefigure[height=3.6cm]{fig0a.eps}
%\titlefigure[height=3.8cm]{fig0a.eps}
%\titlefigurecaption{%
%  Schematic illustration of an etched graphene quantum dot with
%  rough edges. The devices studied are still significantly larger than
%  the presented one.}

\maketitle   % please do not remove

\section{Introduction}

Graphene nanostructures~\cite{han07,che07,dai08,sta08a,pon08,gue08,sch09,gue09,tod09,liu09,rit09} attract
increasing attention mainly due to potential applications in high mobility electronics~\cite{gei07,kat07} and solid state quantum information processing~\cite{tra07}.
In particular, low nuclear spin concentrations expected in graphene promise long spin lifetimes~\cite{tra07,kan05,min06,hue06} and make graphene quantum dots (QDs)~\cite{sta08a,pon08,gue08,sch09,gue09} interesting 
for spin-qubit operations~\cite{tra07}. 
%From a more fundamental point of view 
%On the other hand 
Moreover,
%low-dimensional 
 graphene
nanostructures may %also
 allow to investigate %investigating %(are potentially interesting to investigate)
phenomena related to massless Dirac Fermions in confined dimensions~\cite{pon08,ber87,sch08,rec09,lib09,you09}.  
%quantum electrodynamic (QED) specific phenomena in confined solid state % systems 
%environment~\cite{pon08,ber87,sch08,rec09,lib09,you09}. %pon08,
Intensive research has been triggered by the unique electronic properties of graphene~\cite{cas08} including the gapless linear dispersion, and the unique Landau level (LL) spectrum~\cite{nov05,zha05}. % in this direction.
The search for signatures of graphene-specific properties in % graphene %QDs 
quantum dots is of interest 
in order to understand
%due to their contribution to the magnetic field (B-field) dependence of 
the addition spectra, the spin states and dynamics of confined graphene quasi-particles. %level occupation in general.
 Recent advances in fabricating width-modulated graphene nanoribbons
helped to overcome intrinsic difficulties in (i) creating tunneling barriers and (ii) confining electrons in graphene,
where transport is dominated by Klein tunneling-related phenomena~\cite{dom99,kat06}.
Graphene quantum dots have been fabricated and 
Coulomb blockade~\cite{sta08a,pon08}, quantum confinement%effects
~\cite{sch09,gue09} and charge detection~\cite{gue08} have been observed.
%Challenges of confining electrons in graphene, mainly due to its gap-less band structure and phenomena related to Klein tunneling~\cite{dom99,kat06}
%have been recently taken by width-modulating graphene nanoribbons~\cite{sta08}. Consequently
%Coulomb blockade~\cite{sta08a,pon08}, quantum confinement effects~\cite{sch09} and time-resolved charge detection~[] in graphene QDs have
%been demonstrated recently.

Here, we show tunneling spectroscopy (i.e. transport) measurements on a fully tunable graphene 
quantum dot. %specific signatures in the % B-field dependence of the 
%addition spectrum of a graphene QD.
We present the evolution of a large number of Coulomb resonances near the charge neutrality point in a magnetic field from the low-field regime to the regime of Landau level formation.
In particular, we investigate the quantum dot spectrum in the vicinity of the charge neutrality point as a function of
a perpendicular magnetic field (B-field). Near the electron-hole crossover we observe a rich excited state spectrum with 
inelastic cotunneling lines and unconventional features inside the Coulomb diamonds. Some of these features, which are very pronounced at $B=0$, quickly disappear in a perpendicular magnetic field. Moreover, we find indications of the formation of the lowest ($E=0$) Landau
level at high B-fields marking the electron-hole crossover in graphene devices. 
 
% The addition spectrum displays several intricate features specific to graphene, among them} formation of the lowest ($n=0$) Landau level at high B-fields marking the electron-hole
%crossover.

This paper is organized as follows. In section 2 we briefly review the graphene quantum dot device fabrication.
Transport measurements at zero magnetic field are discussed in section 3.1. Landau level formation and the magnetic field dependence of inelastic cotunneling processes are addressed in sections 3.2 and 3.3, respectively.  

\begin{figure}[t]
\centering % draft=false,keepaspectratio=true,clip,%      width=1.0\linewidth
\includegraphics*[keepaspectratio=true,clip,width=0.9\linewidth]{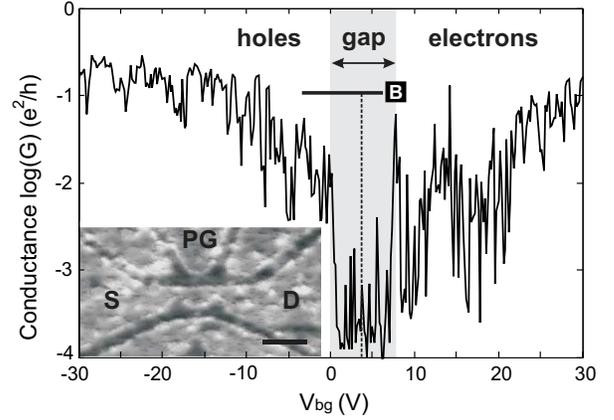}
\caption{
Low bias ($V_{b}=100~\mu$V) back gate characteristic at $V_{pg}=2$~V. The resolved
transport gap (see highlighted region) separates between hole and electron transport.
This trace is a back gate cross-section of Fig.~1(e) in Ref.~\cite{gue09}, which includes the region [B] marked therein.
The dashed line marks the position of the electron-hole crossover identified in Ref.~\cite{gue09}.
The inset shows a scanning force microscope image of an etched graphene quantum dot device with
source (S) and drain (D) leads and a plunger gate (PG) for electrostatic tunability. The scale bar corresponds to 100~nm.
 }
\label{onecolumnfigure1}
\end{figure}

\section{Device fabrication}

The state-of-the-art fabrication process of graphene nanodevices, which has been
mostly developed in Manchester~\cite{nov05,nov04}, and in New York~\cite{zha05}, is 
based on the mechanical exfoliation of (natural) graphite by adhesive tapes~\cite{nov04}.
The substrate material consists of highly
doped silicon (Si$^{++}$) bulk material covered with approximately 300~nm of silicon oxide
(SiO$_2$), where thickness and roughness of the SiO$_2$ top layer is crucial for the
identification and further processing of single-layer graphene samples. Standard photolithography followed by metalization (chrome/gold) and lift-off is used to pattern arrays of reference alignment markers on
the substrate that are later used to re-identify locations (of individual graphene
flakes) on the chip and to align further processing patterns. % including the metallic contacts connecting the flake. Optical microscope, Raman spectroscopy~\cite{fer06,dav07,gupta} and AFM techniques are 
%used to identify and fully characterizing the graphene flakes.

\begin{figure}[t]\centering % draft=false,keepaspectratio=true,clip,%      width=1.0\linewidth
\includegraphics*[keepaspectratio=true,clip,width=0.9\linewidth]{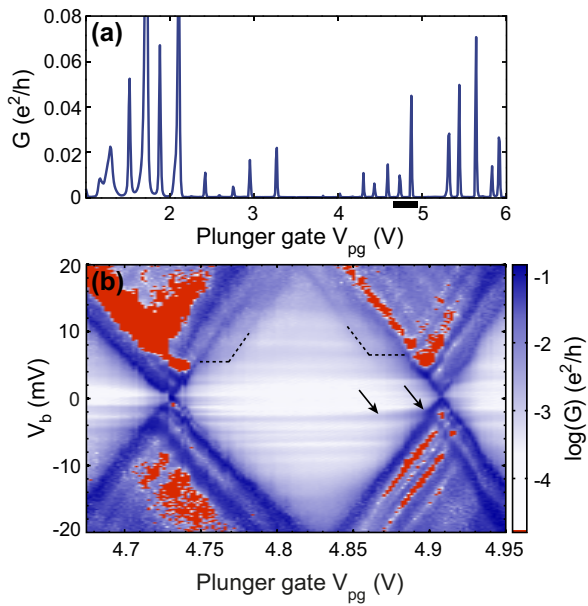}
\caption{%
(color online)
(a) Coulomb blockade resonances of the graphene
quantum dot as function of plunger gate voltage $V_{pg}$ at $V_{bg} = -0.9$~V.
The conductance has been measured at constant bias of $V_b = 100~\mu$V.
(b) Coulomb diamonds, i.e., finite-bias measurement of the quantum dot's differential conductance $G_{qd}=dI/dV_b$ 
as function of $V_b$ and $V_{pg}$ at $B=0$. 
Note the rich excited state spectra and in particular the different vertical
distances between the diamond edge and the diagonal
structures inside the Coulomb-blockaded region for positive $V_{b}$.
}
\label{onecolumnfigure2}
\end{figure}

The graphene flake has to be structured to submicron dimensions in order to
fulfill the device design requirements (see e.g., figure in the abstract).   
We use a technique based on resist spin coating, electron beam lithography (EBL), development and subsequent etching of the unprotected graphene. 
Two successive processing steps have been used including small modifications to decrease the minimum feature size of the
graphene nanodevice. First we used a resist [Polymethylmethacrylat (PMMA)] with a thickness of 45 nm and short etching time to define the delicate structures. In the second step a 100 nm thick resist is used for the coarse structuring of the flake also providing a broader process parameter window for the etching step. % This limits on one hand the minimum feature size of the final graphene devices, and on the other hand the process window for the actual etching step. 
It has been found that short (5 and 15 seconds) mainly physical reactive ion etching (RIE) based on argon and oxygen (80/20) provides good results without influencing the overall quality of the graphene flake~\cite{mol07}. 
%
%With reactive ion etching, the chemical bonds of the etch target are broken up by physical bombardment with argon ions which were created in a plasma. On the bombarded sites, chemical etching (in our case with oxygen) can take place enhanced by the heightened chemical reactiveness of the etchant species due to the plasma. Hence this process combines the anisotropic etching possibilities of the physical bombardment, with the material selectivity of the chemical etching. The ability of this process to etch up to around five-layer thick flakes facilitates contacting structures because it reduces the risk of shorted contacts due to thicker graphitic regions. 
%
After etching and removing the residual EBL resist, 
%scanning force microscope images (as shown e.g. in Figs. 1b,d) are of importance to prove the quality of the patterned graphene flakes mainly in terms of surface contamination. Here the step height  and the surface roughness of the flake, which should be lower than 0.2~nm rms, are good quality measures for flakes before contacting. Selected 
the graphene nanostructures are contacted by an additional EBL step, followed by metalization and lift-off.
%, as illustrated in Figs.~2f-h. Two layers of EBL resist are used to provide a T-shaped resist profile allowing to pattern metal structures down to a lateral size of around 70~nm. After development 
Here we evaporated 2 nm chrome (Cr) and 40 nm gold (Au) for contacting the graphene quantum dot device. 
%
%In Fig. 1c we show an  SFM image of a contacted graphene nanostructure after a successful lift-off process. This image  also proves the alignment between the graphene nanostructure and the metal electrodes to be sufficiently good. The graphene sample is now ready for wire bonding in a chip carrier, which is generally a crucial step for transport studies on nanodevices.

A scanning force microscope image of the device studied here is shown as an inset in Fig.~1. The $50$~nm wide and $80$~nm long graphene quantum dot is connected to source (S) and drain (D) via two graphene constrictions with a width of 25~nm, both acting as tunneling barriers. 
%The two constrictions open a transport gap and lead to the formation of discrete states on the island as sketched in Fig.~1b. 
The dot and the leads can be further tuned by the highly doped silicon substrate used as a back gate (BG) and
an in-plane graphene plunger gate (PG).
%The dot and the leads are tunable by several in-plane graphene gates placed around the structure (PG, SG1, SG2). For the measurements presented in this paper only the plunger gate (PG) and the back gate (BG) are tuned and the nanoribbon charge detector (CD) is not used. 

\begin{figure*}[t]\centering % draft=false,keepaspectratio=true,clip,%      width=1.0\linewidth
\includegraphics*[keepaspectratio=true,clip,width=0.98\linewidth]{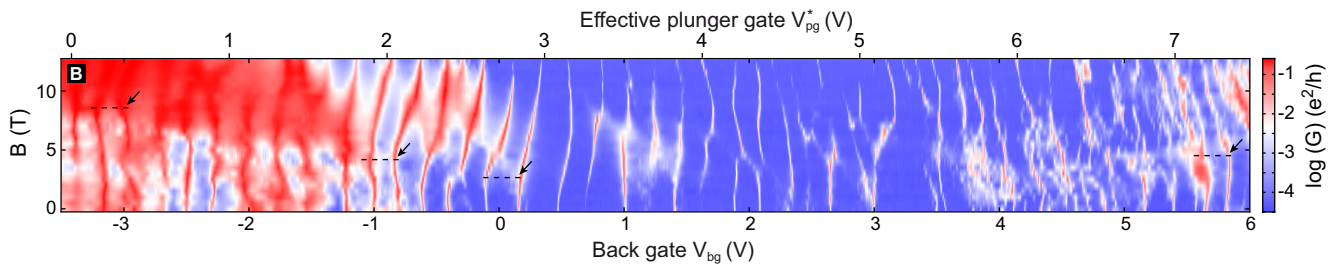}
\caption{%
Evolution of Coulomb blockade resonances in perpendicular magnetic field. The conductance is
plotted as function of back gate voltage $V_{bg}$ and magnetic field B ($V_b$ = 500~$\mu$V). The effective 
plunger gate voltage $V^*_{pg}$, calculated via the relative lever arm, is given on the top of the graph.
This data have been measured in the regime marked by the solid line [B] in Fig.~1(e) of Ref.~\cite{gue09}. }
\label{onecolumnfigure3}
\end{figure*}

\section{Measurements}
All measurements have been performed in a dilution refrigerator
at a base temperature of $T\approx$~90~mK. We have measured the two-terminal conductance
through the graphene quantum dot device by applying a symmetric DC bias
voltage $V_b$ while measuring the current through the quantum dot device with a noise level below 10~fA. For differential conductance measurements a small AC bias,
$V_{b,ac} =$~40~$\mu$V has been superimposed on $V_b$ and the differential conductance
has been measured with lock-in techniques at a frequency of 18.9~Hz. 

In Fig.~1 we show the differential conductance as a function of back gate voltage at low bias ($V_b =$~100~$\mu$V) 
highlighting the strong suppression of the conductance around the charge neutrality point ($0 < V_{bg} < 7.5$~V)
due to the so-called transport gap~\cite{sta09,mol09}. Here we tune transport from
the deep hole to the deep electron regime, as illustrated in Fig.~1.
The large number of resonances within the gap region may be due to both, (i) resonances
in the graphene constrictions acting as tunneling barriers~\cite{sta08a} (and thus being main responsible
for the large extension of this transport gap) and (ii) Coulomb resonances of the quantum dot itself.

\subsection{Coulomb blockade measurements at $B=0$}
By fixing the back gate voltage at a value close to the transport gap (e.g., $V_{bg} = -0.9$~V) and
sweeping the lateral plunger gate voltage ($V_{pg}$) in a narrow range close to the charge neutrality point (see below) Coulomb blockade resonances of the graphene
quantum dot can be well resolved, as shown in Fig.~2(a). 
The strong amplitude modulation of the conductance peaks [see e.g., on the very left of Fig.~2(a)]
is mainly due to transparency modulations of the constrictions~\cite{sta08a}, which can dominate the width of the Coulomb resonances and significantly elevate the conductance. In Fig.~2(b) we show corresponding
Coulomb diamond measurements [see black bar in Fig.~2(a)], that are measurements of the differential conductance ($G = dI/dV_{b}$)
as a function of bias voltage $V_b$ and plunger gate voltage $V_{pg}$. Note, that this measurement was recorded at a higher base temperature of $T=370$~mK. The reason for this slightly higher temperature is the good visibility of the diagonal lines inside the diamond. The overall structure of the diamond is not changed significantly by increasing the temperature. 
The logarithmic plot of the Coulomb diamonds measured in the 
close vicinity of the charge neutrality point displays a rich excited state spectrum. We observe lines of increased conductance 
outside the diamonds running parallel to the diamond edges, which are 
well aligned to inelastic cotunneling onsets visible as faint horizontal (constant bias) structures inside the diamond-shaped
regions (see arrows). 
At the diamond
boundary, the horizontal lines seamlessly join some
of the most prominent diagonal lines in the non-blockaded
region, allowing to extract characteristic excitation energies in the
range of 2 to 3 meV.

In addition, we observe for positive bias $V_{b} > 0$~V unconventional features inside the 
Coulomb-blockaded regions (see dashed lines) consisting of diagonal
lines running parallel to the diamond edge. 
The fact
that they have the same slope as the diamond edges suggests
that they are connected to the alignment of an energy
level with source (negative slope) or drain (positive slope).
The vertical distance
between the diagonal lines and the diamond edge is identical
for lines with positive and negative slope (see dashed line inside the diamond). 
%When extended towards higher or towards negative
%voltages, most of the diagonal lines apparently join prominent
%lines in the nonblockaded regime.
It has been shown by Schleser et al.~\cite{sch05} that these features are related to
sequential tunneling through the quantum dot, occurring after it has been excited by an inelastic cotunneling
process. %According to ref.~\cite{sch05} these processes can be explained in detail using transport calculations within the real-time Green's function approach where diagrams up to fourth order in the tunneling matrix elements are included. 
In simple terms, these features can be attributed to a transport configuration where the tunneling-out rate from the excited state is significantly larger than the relaxation rate to the ground state, which might be directly related to strongly non-linear energy dependences of the 
tunneling barriers.   
Finally, we also observe regions of negative differential conductance (see red areas), which are not yet fully understood. 

\begin{figure*}[htb]%
  \includegraphics*[keepaspectratio=true,clip,width=.67\textwidth]{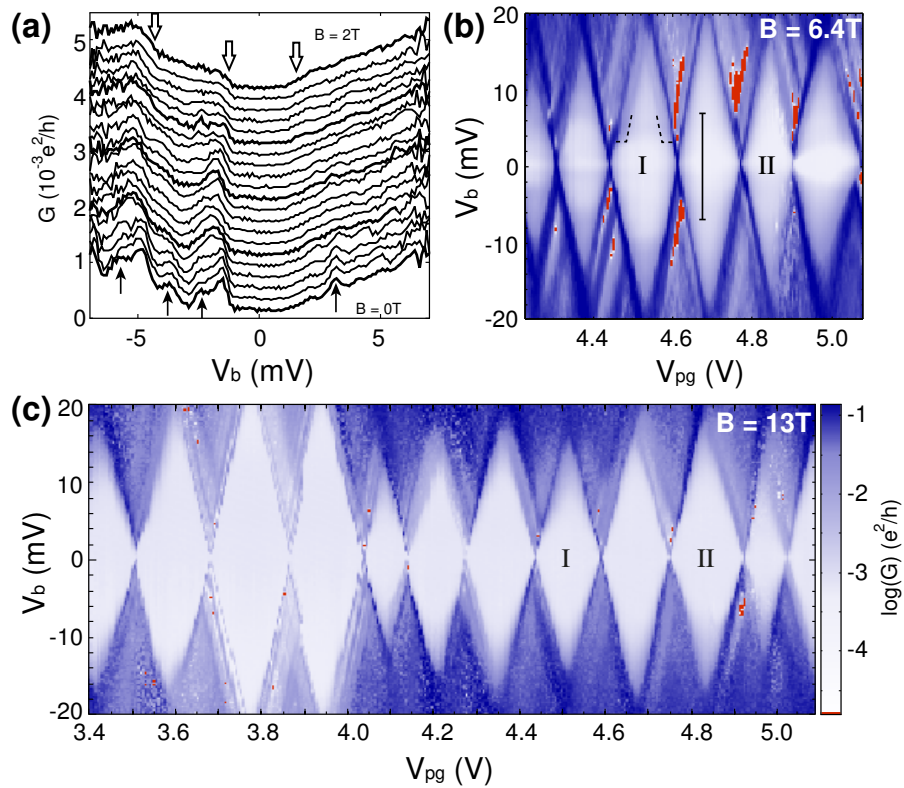}%
  \caption[]{% 
  (color online)
(a) Evolution of inelastic cotunneling onsets inside the
Coulomb-blockaded region in perpendicular magnetic field.
The conductance is plotted as function of bias $V_b$ and
magnetic field B~=~0 to 2~T (stepped in units of 0.1~T) for constant
plunger gate voltage $V^{*}_{pg}~=~$4.66~V [see vertical line in panel (b)].
For clarity the individual traces have been offset by
0.2~$\times 10^{-3} e^2$/h. The peaks related to inelastic cotunneling marked by arrows quickly 
vanish as function of B.
(b,c) Coulomb diamond measurements at B~=~6.4~T (b)
and B~=~13~T (c), respectively. These measurements are taken at constant back gate
voltage $V_{bg}$~-0.9~V and highlight both (i) single dot signatures at high magnetic fields and (ii) 
the strong suppression of excited state-related signatures at higher B-fields.} 
    \label{sidecaption}
\end{figure*}

\subsection{Coulomb resonances as a function of a perpendicular magnetic field}

In Fig.~3 we show more than 40 Coulomb resonances
as a 
function of a B-field perpendicular to the graphene plane.
The measurement has been taken in the back gate voltage range from $V_{bg}$ =
-3.5 to 6~V (also highlighted by the vertical line [B] in Fig.~1 and Fig.~1(e) in Ref.~\cite{gue09}).
Thus we certainly tune
from hole to electron transport (see arrows in Fig.~1). The evolution of Coulomb resonances in (perpendicular)
magnetic field shows signatures of the electron-hole
crossover of the quantum dot states (see below and Ref.~\cite{gue09}). There is a
common trend of resonances at lower back gate voltages (see,
e.g., resonances at $V_{bg} = 0$~V) to bend for increasing B-field
towards higher energies (higher $V_{bg}$). In contrast we find
for higher back gate voltage the opposite trend (see,
e.g., resonances and arrow at $V_{bg}~\approx$~5.8~V), where resonances tend to shift to
lower energies for increasing magnetic field. This overall pattern
is disturbed by additional features such as
localized states, regions of multi-dot behavior and strong
amplitude modulations due to constriction resonances.
For example, we observe a strongly elevated background below
$V_{bg} <$~-1~V, which can be attributed to the increase of 
the transparency of the tunneling barriers in agreement with the measurement
presented in Fig.~1.  Additionally, we observe between $V_{bg}$~=~-1 and 0~V 
a weakly
coupled state crossing the Coulomb resonances at finite magnetic fields.
We interpret the weak
magnetic field dependence and low visibility in transport
as a manifestation of a strongly localized state. Moreover,
we see several level crossings and splittings in
the region above $V_{bg}$~=~4~V, which might be due to the presence of
an additional quantum dot spontaneously forming in one of the two constrictions, and coupling to the gates with different lever arms~\cite{gue09}.

Finally, in close analogy to Ref.~\cite{gue09} where the same sample was investigated in a different parameter regime we observe "kink" features with increasing B-field. It has been found that
with increasing magnetic field the levels feature a kink signifying filling factor $\nu = 2$ in the dot (see dashed lines and arrows in Fig.~3) before converging toward the $E = 0$ Landau level, in analogy to what has been observed in GaAs quantum dots~\cite{cio00}. In particular, the population
of the lowest Landau level leads to a region
around $E = 0$~\cite{gue09,lib09b} with a remarkably ordered addition spectrum, which emerges for large magnetic fields (see also the analytical calculations for a circular
dot~\cite{sch08,rec09}).
This feature is a unique consequence of the interplay of the unique
Landau level spectrum in graphene~\cite{nov05,zha05} and the carrier confinement due to
the finite size of the system.
For more details, including numerical simulations of graphene quantum dots and
a quantitative discussion of the peak-to-peak spacing evolution in B-field
we refer to Refs.~\cite{gue09,lib09b}.

\subsection{Coulomb diamonds at finite magnetic field}

In Fig.~4a we show the evolution of inelastic cotunneling onsets inside the
Coulomb-blockaded region as a function of perpendicular magnetic field.
The conductance is plotted as a function of bias voltage $V_b$ and
magnetic field at constant
plunger gate voltage $V_{pg}~=~$4.66~V [see vertical line in Fig.~4(b)] and
the B-field has been stepped by 0.1~T from B~=~0 to 2~T.
While at B~=~0~T we observe several peaks and steps (black and white arrows), at B~=~2~T the trace is smoother with remaining steps at 1.4~mV, -1.3~mV and 4.5~mV (white arrows). The different B-field dependence of the features might be attributed to two origins. The fine structures disappearing at B~=~0.15~T (black arrows) may arise from energy dependent fluctuations of the barrier transmission, while the steps (white arrows) are attributed to the onset of inelastic cotunneling channels.
%%CS version
%The peaks related to inelastic cotunneling, well visible for low magnetic fields (see black
%arrows at -5, -3.9, -2.4 and 3.2~mV) quickly 
%vanish for increasing B-field. Consequently, at B~=~0.15~T the fine structures at low bias voltages completely disappear and it strongly suggests that inelastic cotunneling processes in this system are suppressed at 
%finite magnetic field.
This is supported by Figs.~4(b),(c), where we show Coulomb diamond measurements at  
B~=~6.4~T and B~=~13~T. These measurements are taken at the same constant back gate
voltage $V_{bg} =$~-0.9~V as in Fig.~2(b) and are plotted with the same colorscale.
Thus, we can directly compare Figs.~2(b), 4(b) and 4(c) simplified by the diamond labels I and II [see Figs.~4(b),(c); the diamond shown in Fig.~2(b) corresponds to diamond II]. 
The horizontal (constant bias) lines in the Coulomb blockaded region present at B~=~0~T are absent at higher B-fields. However, at B~=~6.4~T cotunneling onsets and the unconventional feature consisting of diagonal lines
inside the diamond [see above and dashed lines in Fig.~4(b)] can still be observed, but in a different diamond than at B~=~0~T (I instead of II). At B~=~13~T all excited-state related signatures inside the diamond are suppressed.
In addition, also the negative differential conductance outside the diamond is significantly reduced and might also be related to a decrease in the energy dependent fluctuation of the barrier transmission.
However, from these measurements we can conclude that the single-dot behavior of this etched graphene quantum dot device stays well preserved also for high magnetic fields.

\section{Conclusion}
We have performed detailed studies of transport through a graphene quantum dot in the vicinity of the charge neutrality point. The evolution of Coulomb resonances in magnetic field showed signatures of Landau level formation. Indications for the crossing of filling factor $\nu = 2$ are obtained by the observation of "kinks" in spectral lines before bending towards the charge neutrality point. 
Coulomb blockade diamonds at $B = 0$~T near the electron-hole crossover show many excited states and cotunneling onsets accompanied by diagonal lines in the diamond attributed to cotunneling mediated sequential tunneling through an excited state. Many small scale lines inside the diamond disappear quickly ($B = 1.5$~T) in a magnetic field while more step like features (cotunneling onsets) are still visible at $B = 6.4$~T. The presented measurements open the way for an in-depth exploration of the few charge carrier regime including addition spectra in graphene quantum dots.

%Although not fully understood in detail these measurements present a towards  measurements are not fully understood in detail.
%
%-Coulomb blockade at B=0T: features in diamond, cotunneling mediated sequential tunneling (co-set)
%-
%-Coulomb diamonds at magnetic fields: two origins of differential conductance changes in diamond: 
%(i) energy dependent conductance fluctuations (rather peaks than steps) with fast B-decrease
%(ii) co-tunneling onsets with slower B-dependence
%
%We discussed the spectra of excited states of a graphene quantum dot in a perpendicular magnetic field. 
%
%
%% by tuning the Fermi level from electrons to holes. 
%%Despite of several  charging effects
%We observed signatures for the development of the graphene specific $E = 0$ Landau level around the charge neutrality point. The overall tendency of states converging towards the zero-energy Landau level was accompanied by kinks which may be attributed to particular filling factors in the quantum dot. Coulomb diamond measurements in this transition regime show a rich spectrum of excited states and cotunneling. %Although no systematic filling of electrons is found so far 
%These measurements open the way for more involved studies of the electron-hole transition including a better understanding of addition spectra, and spin states in graphene quantum dots.

%\begin{acknowledgement}
The authors wish to thank F.~Libisch, J.~Seif, P.~Studerus, C.~Barengo, T.~Helbling and S.~Schnez for help and discussions. 
Support by the ETH FIRST Lab, the Swiss National Science Foundation and NCCR nanoscience are gratefully acknowledged. 
%\end{acknowledgement}

% Use the following code if you wish to generate your bibliography with BibTeX;
% replace the string "pss-demo" below with the name(s) of
% the BibTeX data base(s) you want to use.
% The resulting bibliography-output (the contents of the .bbl file)
% must be pasted back into this file before submission.
%
% \bibliographystyle{pss}
% \bibliography{pss-demo}
%
% Replace the following example bibliography with your references
% before submission:

\end{document}